\documentstyle[12pt]{article}

\def\hybrid{\topmargin -20pt  \oddsidemargin 0pt
      \headheight 0pt   \headsep 0pt
      \textwidth 6.25in % A4 paper
      \textheight 9.5in % A4 paper
      \marginparwidth .875in
      \parskip 5pt plus 1pt   \jot = 1.5ex}

\hybrid

\let\LARGE=\large

\let\large=\normalsize

\begin{document}
%\titlepage
\def\x{\times}
\def\beq{\begin{equation}}
\def\eeq{\end{equation}}
\def\beqa{\begin{eqnarray}}
\def\eeqa{\end{eqnarray}}
\def\L{ {\cal L}}
\def\C{ {\cal C}}
\def\N{ {\cal N}}
\def\calE{{\cal E}}
\def\lin{{\rm lin}}
\def\Tr{{\rm Tr}}
\def\cF{{\cal F}}
\def\cD{{\cal D}}
\def\modS{{S+\bar S}}
\def\mods{{s+\bar s}}
\newcommand{\Fg}[1]{{F}^{({#1})}}
\newcommand{\cFg}[1]{{\cal F}^{({#1})}}
\newcommand{\cFgc}[1]{{\cal F}^{({#1})\,{\rm cov}}}
\newcommand{\Fgc}[1]{{F}^{({#1})\,{\rm cov}}}
\def\mpl{m_{\rm Planck}}
\def\mxth{\mathsurround=0pt }
\def\xversim#1#2{\lower2.pt\vbox{\baselineskip0pt \lineskip-.5pt
x  \ialign{$\mxth#1\hfil##\hfil$\crcr#2\crcr\sim\crcr}}}
\def\simgr{\mathrel{\mathpalette\xversim >}}
\def\simle{\mathrel{\mathpalette\xversim <}}

\newcommand{\ms}[1]{\mbox{\scriptsize #1}}
\renewcommand{\a}{\alpha}
\renewcommand{\b}{\beta}
\renewcommand{\c}{\gamma}
\renewcommand{\d}{\delta}
\newcommand{\th}{\theta}
\newcommand{\TH}{\Theta}
\newcommand{\pa}{\partial}
\newcommand{\g}{\gamma}
\newcommand{\G}{\Gamma}
\newcommand{\A}{\Alpha}
\newcommand{\B}{\Beta}
\newcommand{\D}{\Delta}
\newcommand{\e}{\epsilon}
\newcommand{\E}{\Epsilon}
\newcommand{\z}{\zeta}
\newcommand{\Z}{\Zeta}
\newcommand{\k}{\kappa}
\newcommand{\K}{\Kappa}
\renewcommand{\l}{\lambda}
\renewcommand{\L}{\Lambda}
\newcommand{\m}{\mu}
\newcommand{\M}{\Mu}
\newcommand{\n}{\nu}
\newcommand{\X}{\Chi}
\newcommand{\R}{\Rho}
\newcommand{\s}{\sigma}
\renewcommand{\S}{\Sigma}
\renewcommand{\t}{\tau}
\newcommand{\T}{\Tau}
\newcommand{\y}{\upsilon}
\newcommand{\Y}{\upsilon}
\renewcommand{\o}{\omega}
\newcommand{\q}{\theta}
\newcommand{\h}{\eta}

\def\dota{ {\dot{\alpha}} }
\def\lag{Lagrangian}
\def\Kahler{K\"{a}hler}
\def\kahler{K\"{a}hler}
\def\A{ {\cal A}}
\def\C{ {\cal C}}
\def\F{{\cal F}}
\def\cL{ {\cal L}}

\def\R{ {\cal R}}
\def\x{ \times }
\def\beq{\begin{equation}}
\def\eeq{\end{equation}}
\def\beqa{\begin{eqnarray}}
\def\eeqa{\end{eqnarray}}

\sloppy
\newcommand{\ba}{\begin{array}}
\newcommand{\ea}{\end{array}}
\newcommand{\be}{\begin{equation}}
\newcommand{\eq}{\end{equation}}
\newcommand{\ov}{\overline}
\newcommand{\un}{\underline}
\newcommand{\p}{\partial}
\newcommand{\la}{\langle}
\newcommand{\ra}{\rangle}
\newcommand{\bl}{\boldmath}
\newcommand{\ds}{\displaystyle}
\newcommand{\nl}{\newline}
\newcommand{\Nzahl}{{\bf N}  }
\newcommand{\zzahl}{ {\bf Z} }
\newcommand{\Zzahl}{ {\bf Z} }
\newcommand{\Qzahl}{ {\bf Q}  }
\newcommand{\Rzahl}{ {\bf R} }
\newcommand{\Czahl}{ {\bf C} }
\newcommand{\wt}{\widetilde}
\newcommand{\wh}{\widehat}
\newcommand{\fs}[1]{\mbox{\scriptsize \bf #1}}
\newcommand{\ft}[1]{\mbox{\tiny \bf #1}}
\newtheorem{satz}{Satz}[section]
\newenvironment{Satz}{\begin{satz} \sf}{\end{satz}}
\newtheorem{definition}{Definition}[section]
\newenvironment{Definition}{\begin{definition} \rm}{\end{definition}}
\newtheorem{bem}{Bemerkung}
\newenvironment{Bem}{\begin{bem} \rm}{\end{bem}}
\newtheorem{bsp}{Beispiel}
\newenvironment{Bsp}{\begin{bsp} \rm}{\end{bsp}}
\renewcommand{\arraystretch}{1.5}

%\textwidth14.5cm
%\textheight23.0cm
%\oddsidemargin0.5cm
%\topmargin-1.4cm

%\addtocounter{section}{1}

\renewcommand{\thesection}{\arabic{section}}
\renewcommand{\theequation}{\thesection.\arabic{equation}}

%\setcounter{section}{1}
%\addtocounter{section}{1}
\parindent0em

\def\S4{\frac{SO(4,2)}{SO(4) \otimes SO(2)}}
\def\P3{\frac{SO(3,2)}{SO(3) \otimes SO(2)}}
\def\MGd{\frac{SO(r,p)}{SO(r) \otimes SO(p)}}
\def\SOd{\frac{SO(r,2)}{SO(r) \otimes SO(2)}}
\def\SO2{\frac{SO(2,2)}{SO(2) \otimes SO(2)}}
\def\SUm{\frac{SU(n,m)}{SU(n) \otimes SU(m) \otimes U(1)}}
\def\SUS{\frac{SU(n,1)}{SU(n) \otimes U(1)}}
\def\SK{\frac{SU(2,1)}{SU(2) \otimes U(1)}}
\def\SU{\frac{ SU(1,1)}{U(1)}}

\begin{titlepage}
\begin{center}
\hfill CERN-TH/97-61\\
\hfill HUB-EP-97/23\\
\hfill UPR-746-T\\
\hfill {\tt hep-th/9704095}\\

\vskip .3in

{ \LARGE \bf Quantum N = 2 Supersymmetric Black Holes in the $S$-$T$ Model}

\vskip .3in

{\bf Klaus Behrndt$^a$, Gabriel Lopes Cardoso$^b$ and Ingo Gaida$^c$
}\footnote{\mbox{email: \tt 
behrndt@qft2.physik.hu-berlin.de,
cardoso@afsmail.cern.ch, } \\ \tt 
gaida@cvetic.hep.upenn.edu} 
\\
%\vskip 1.2cm
\vskip 1cm

$^a${\em Humboldt-Universit\"at zu Berlin,
Institut f\"ur Physik, 
D-10115 Berlin, Germany}\\
$^b${\em Theory Division, CERN, CH-1211 Geneva 23, Switzerland}\\
$^c${\em Department of Physics and Astronomy,
University of Pennsylvania, \\
Philadelphia, PA 19104-6396, USA}

\vskip .1in

\end{center}

\vskip .2in

\begin{center} {\bf ABSTRACT } \end{center}
\begin{quotation}\noindent
We consider axion-free quantum corrected black hole solutions in the
context of the heterotic $S$-$T$ model with half the $N=2, D=4$
supersymmetries unbroken.  We express the perturbatively corrected
entropy in terms of the electric and magnetic charges in such a way, that
target--space duality invariance is manifest.  
We also discuss the microscopic
origin of particular quantum black hole configurations. We propose a
microscopic interpretation in terms of a gas of closed membranes for
the instanton corrections to the entropy.
\end{quotation}
\vskip .5cm
% April 1997\\
CERN-TH/97-61\\
\hfill April 1997\\
\end{titlepage}
\vfill
\eject

%%%%%%%%%%%%%%%%%%%%%%%%%%%%%%%%%%%%%%%%%%%%%%%%%%%%%%%%%%%%%
\newpage

\section{Introduction}

Black holes play an important role in string theory, and in recent times
there has been considerable progress in the understanding of microscopic
and macroscopic properties of supersymmetric black holes in string theory
(for a review see \cite{review}).

It is well known from classical general relativity that non-rotating
black hole solutions can be parametrized in terms of electric and
magnetic charges and the ADM mass, only. In the context of string
theory, it has been shown in \cite{la/wi} that four-dimensional
non-rotating black hole solutions in the BPS limit depend classically
only on the bare quantized charges on the horizon. Thus, the black
hole solutions in the BPS limit are, on the horizon, independent 
of the values of the
moduli at spatial infinity.  In \cite{FerKal1} it has been shown how
one can understand this result from a supersymmetric point of view: On
the horizon the central charge of the extended supersymmetry algebra
acquires a minimal value and thus the extremization of the central
charge provides the specific moduli values on the horizon
\cite{FerKal1,FerGiKal}.  Moreover, the entropy of certain
supersymmetric black hole configurations can depend on additional
topological data. In the context of a Calabi--Yau compactification
these can be, for instance, the intersection numbers, the second Chern
class and the Euler number \cite{BCDKLM}.

Although the BPS limit of black hole solutions in four dimensions with
$N \geq 4$ is by now well understood \cite{D4N4Entr}, new features of
black hole physics arise in four-dimensional $N=2$ string theory.  In
particular there exists a large number of different $N=2$ string vacua
so that the extreme black hole solutions depend on the specific
details of the particular $N=2$ string model. Consequently the same
features are present for the $N=2$ entropy formula. 
% /////
% Nevertheless, on the horizon the $N=2$ entropy depends only on
% conserved quantities, i.e.\ classically on the quantized magnetic and
% electric charges only, although the nature of dependence is governed
% by the particular string model. //////

The $N=2$ central charge and the $N=2$ BPS mass spectrum can be directly 
calculated form the $N=2$ holomorphic prepotential. Therefore the
parameters of the prepotential of a given $N=2$ string model
determine the black hole entropy as well as the values of the scalar
fields on the horizon.

If one considers four-dimensional $N=2$ heterotic string compactifications
on $K3 \times T_{2}$ with $N_{V}+1$ vector multiplets (including the
graviphoton), the classical prepotential is completely universal and
corresponds to a scalar non-linear $\sigma$-model based on the coset space
$\frac{SU(1,1)}{U(1)} \otimes \frac{SO(2,N_{V}-1)}{SO(2) \times SO(N_{V}-1)}$.
The corresponding classical $N=2$ black hole entropy and the moduli
on the horizon have been computed explicitly in \cite{BKRSW,CLM}.

Since in heterotic $N=2$ string compactifications the dilaton can
be described by a vector multiplet, the heterotic prepotential
receives perturbative quantum corrections only at the one-loop level
\cite{DKLL,AFGNT}; in addition there are non-perturbative contributions.
The perturbative (and non-perturbative) corrections generically split
into a cubic polynomial, a constant term and an infinite series of 
polylogarithmic terms.
Thus, quantum black hole solutions are generically determined by an infinite 
set of integer numbers. Hence, the extremization problem of the quantum
corrected $N=2$ central charge is, in general, difficult to solve. 
Nevertheless,
we will be able to give explicit examples, where all the perturbative quantum
corrections are taken into account, and where the extremization problem can
still be solved completely.

In \cite{BCDKLM, Rey} a simple implicit formula for the black hole entropy
in terms of the heterotic string coupling and the target--space duality
invariant inner product of charges has been given, which holds to all
orders in perturbation theory. This result is the starting point of
the present paper and we will discuss it in the context of the
heterotic $S$-$T$ model. 
%Related questions have been recently
%considered in \cite{Rey}.  

The paper is organized a follows: In the first section we will briefly
introduce the $N=2$ vector couplings, the $N=2$ central charge and the
related Bekenstein-Hawking entropy in terms of the $N=2$ prepotential.
In section 3 we introduce the heterotic $S$-$T$ model, its perturbative
and non-perturbative quantum corrections and the corresponding
transformation laws under perturbative target--space duality. In section
4 we discuss axion-free black holes in the $S$-$T$ model. We treat
most of the cases explicitly in terms of target--space duality invariant
combinations of quantized charges. In one case, we also discuss the implicit
axion-free black hole entropy in the $S$-$T$ model including all 
perturbative and non-perturbative quantum corrections. Then we solve
this case for a special weak coupling limit near the line  of
gauge symmetry enhancement $S=T$ in moduli space.
Section 5 is
devoted to the 10 and 11 dimensional configurations that yield
the black hole solutions upon compactification. Finally in section
6 we propose a microscopic interpretation for the Bekenstein-Hawking
entropy in terms of an intersection of $M$-branes living in a gas
of closed membranes. In the last section we summarize our results.

%%%%%%%%%%%%%%%%%%%%%%%%%%%%%%%%%%%%%%%%%%%%%%%%%%%%%%%%%%%%%%%%%%%%

\section{N = 2 supergravity and special geometry}

\setcounter{equation}{0}

The vector couplings of $N=2$ supersymmetric Yang-Mills theory
are encoded in a holomorphic function $F(X)$, where $X$ 
denotes the complex scalar fields of the  
vector supermultiplets. With local supersymmetry this function 
depends on one extra field, in order to incorporate the 
graviphoton. The theory can then be encoded in terms of a 
holomorphic function $F(X)$ which is homogeneous of second 
degree and depends on complex fields $X^I$ with $I=0,1,\ldots 
N_V$. Here $N_V$ counts the number of physical vector multiplets. 

The resulting {\em special geometry} \cite{DWVP,special} can be defined 
more abstractly in  
terms of a symplectic section $V$, also referred to as period 
vector: a $(2N_V+2)$-dimensional 
complex symplectic vector, expressed in terms of the holomorphic 
prepotential $F$ according to 
\beqa
V= \pmatrix{ X^I \cr F_J\cr} \;\;, \label{sympsection}
\eeqa
where $F_I=\partial F/\partial X^I$. The $N_V$ physical scalar fields
of this system parametrize 
an $N_V$-dimensional
complex hypersurface, defined by the condition that the section satisfies a 
constraint 
\beqa
\langle \bar V,V\rangle \equiv \bar V^{\rm T}\Omega V 
= -i ,\label{symconstr}
\eeqa 
with $\Omega$ the antisymmetric matrix
\beqa
\Omega = \pmatrix{ 0& {\bf 1} \cr -{\bf 1} &0 \cr }\,.
\eeqa
The embedding of this hypersurface can be described in terms of
$N_V$ complex coordinates $z^A$ 
($A=1,\dots ,N_V$) by letting the $X^I$ be proportional to
some holomorphic sections $X^I(z)$ of the complex projective space.
In terms of these 
sections the $X^I$ read
\beq
X^I = e^{{1\over 2}K(z,\bar z)}\,
X^I(z)\,,\label{section} 
\eeq
where $K(z,\bar z)$ is the K\"ahler potential, to be introduced 
below. 
In order to distinguish the sections $X^I(z)$ from the original quantities
$X^I$, we will always explicitly indicate their $z$-dependence.
The overall factor $\exp[{1\over2}K]$ is chosen such that the constraint
(\ref{symconstr}) is satisfied. Furthermore, by virtue of the 
homogeneity property of $F(X)$, we can extract an overall factor 
$\exp[{1\over 2}K]$ from the symplectic sections (\ref{sympsection}), so 
that we are left with a holomorphic symplectic section. Clearly 
this 
holomorphic section is only defined projectively, i.e. modulo 
multiplication by an arbitrary holomorphic function. On the 
K\"ahler potential these projective transformations act as 
K\"ahler transformations, while on the sections $V$ they act as 
phase transformations. 
 
The resulting geometry for the space of physical scalar fields
belonging to vector multiplets of an $N=2$ 
supergravity theory is a special K\"ahler geometry,
with a K\"ahler metric $g_{A\bar B}=\partial_A\partial_{\bar B}K(z,\bar z)$
following from a K\"ahler potential of the special 
form
\begin{equation}
K(z,\bar z)=
-\log\Big(i \bar X^I(\bar z)F_I ( X^I(z))-i X^I(z)
\bar F_I(\bar X^I(\bar z))\Big) .
\label{KP} 
\end{equation}
A convenient choice of inhomogeneous coordinates $z^A$
are the {\em special} coordinates, defined by 
\begin{equation}
X^0(z)=1\,,\quad X^A(z)= z^A\,,\qquad A=1,\ldots ,N_V.
\end{equation}
In this parameterization the K\"ahler potential can be written as 
\cite{SU} 
\begin{equation}
K(z,\bar z) = -\log\Big(2({\cal F}+ \bar{\cal F})-
           (z^A-\bar z^A)({\cal F}_A-\bar{\cal F}_A)\Big)\,,
\label{Kspecial}
\end{equation}
where ${\cal F}(z)=i(X^0)^{-2}F(X)$. 

We should point out that it is possible to rotate the basis 
specified by (\ref{sympsection}) by an $Sp(2N_V+2,{\bf Z})$ 
transformation in such a way that it is no longer possible to 
associate them to a holomorphic function \cite{Ceresole}. 
The supergravity  
Lagrangian is then expressed entirely in terms of the symplectic 
section $V = (P^I, i Q_J)^T$, 
without restricting its parametrization so as to  
correspond to a prepotential $F(X)$ \cite{Ceresole}.

The target--space duality group $\Gamma$ is a certain subgroup
of $Sp(2N_V+2,{\bf Z})$. 
Under target--space duality transformations, 
the period vector $V$ transforms as a symplectic vector:
\begin{eqnarray}
\tilde X{}^I&=&U^I_{\ J}\,X^J + Z^{IJ}\,
F_J\, ,\hspace{2cm}
\tilde F{}_I \ = \ V_I{}^J\,F_J + W_{IJ}\,X^J \,,
\label{xxx}
\end{eqnarray}
where $U$, $V$, $W$ and $Z$ are constant, real,  $(N_V+1)\times(N_V+1)$
matrices, which have to satisfy the symplectic constraint
\begin{equation}
{\cal O}^{-1} = \Omega\, {\cal O}^{\rm T} \,\Omega^{-1} \qquad
\mbox{where} \quad {\cal O} =
\pmatrix{ U & Z \cr W  & V\cr }\;.
\label{spc}
\end{equation}

Finally consider $N=2$ BPS states, whose masses are equal
to the central charge $Z$ of the $N=2$ supersymmetry algebra.
In terms of the magnetic/electric charges $(p^I,q_J)$ 
and the period vector $V=(X^I,F_J)^T$
the BPS masses take the following form \cite{Ceresole}:
\beqa
M_{BPS}^2=|Z|^2=e^{K(z,\bar z)}|q_IX^I(z)-p^IF_I(z)|^2
=e^{K(z,\bar z)}\,|{\cal M}(z)|^2.\label{bpsmasses1}
\eeqa
It follows that $M_{BPS}^2$ is invariant under symplectic transformations
(\ref{xxx}).

In the symplectic basis where the symplectic section $V$ is given by 
$V = (P^I, i Q_J)^T$,  the BPS mass takes the following form \cite{Ceresole}:
\beqa
M_{BPS}^2=|Z|^2=e^{K(z,\bar z)}|M_I P^I(z) + i N^I Q_I(z)|^2
=e^{K(z,\bar z)}\,|{\cal M}(z)|^2.\label{bpsmasses2}
\eeqa
We will choose $V = (P^I, i Q_J)^T$ in such a way that 
the symplectic quantum numbers $(N^I,M_J)$ and the 
charges $(p^I, q_J)$ are related as follows,
\beqa
N^I &=& (p^0, q_1, p^2, \dots, p^{N_V}), \hspace{1cm}
M_J \ = \ ( q_0, - p^1, q_2, \dots, q_{N_V}) \;\;\;.
\eeqa
The BPS mass formula (\ref{bpsmasses1}), when evalutated on the horizon
of a BPS black hole, also yields its entropy.  On the horizon, 
the moduli fields take their fixed values, and these fixed values can be
determined by solving a set of  $2N_{V}+2$
extremisation conditions \cite{FerKal1}.  
In a suitable basis $Y$, given by
$Y^I = {\bar Z} X^I$ \cite{BCDKLM},
these $2N_{V}+2$ extremisation equations are then given by
\begin{eqnarray}
Y^{I}- \bar Y^{I} &=& i p^{I}, \hspace{2cm}
F_{I}- \bar F_{I} \ = \  i q_{I}\;\;\;,
\label{extremcon}
\end{eqnarray}
and the 
Bekenstein-Hawking entropy reads
\begin{eqnarray}
\label{bh_ent}
S_{BH} &=&  \, \pi \, |Z|_{|{\rm fix}}^{2} 
= \pi \ \left( |Y^0|^2 e^{-K(z, {\bar z})} \right)_{\rm | fix} 
= i \pi \ \left( 
\bar Y^{I}  F_{I}( Y^{I}) 
-  Y^{I} \bar F_{I} (\bar Y^{I}) 
\right)_{| {\rm fix}} \;.
\end{eqnarray}
These expressions are valid on the horizon or as double extreme
black holes \cite{FerKal1}. For a discussion of more general
black holes, where one replaces the charges by harmonic functions, 
see \cite{Sabra}.

%%%%%%%%%%%%%%%%%%%%%%%%%%%%%%%%%%%%%%%%%%%%%%%%%%%%%%%%%%%%%%%%%%
\section{The $S$-$T$ model}

\setcounter{equation}{0}

\subsection{General formulae}

In the following, we will focus on the two parameter model 
\cite{KV,KLM,KKLMV,HenMoo}
based on a type IIA compactification on a Calabi--Yau space given by
a degree $12$ hypersurface in the weighted projective space
${\bf P}^4_{(1,1,2,2,6)}$ with Hodge numbers
$(h^{1,1},h^{2,1})= (2,128)$ and Euler number $\chi = 2(
h^{1,1} - h^{2,1}) = - 252$.  
On the type II side, the vector multiplet prepotential is given by
\cite{HoKle,KKLMV,HenMoo}
\beqa
{\cal F}_{II} &=& - t_1 (t_2)^2 - \frac{2}{3} (t_2)^3 - c 
+ \frac{1}{8 \pi^3} \sum_{j \geq 0, k \geq 1} n_{k,j}
Li_3(e^{-2\pi(jt_1+kt_2)}) \nonumber\\
& & + 
 \frac{1}{8 \pi^3} n_{0,1}
Li_3(e^{-2\pi t_1}) \;\;\;,
\label{preptype2}
\eeqa
where $c = \frac{ \chi \zeta(3)}{16 \pi^3}$.
Here, 
$t^1 = i z^1$ and $t^2 = i z^2$ denote the two coordinates
of the K\"ahler cone.  
The instanton numbers 
$n_{k,j}$ can be found in \cite{HoKle,HenMoo}.  Note that $ n_{0,1} = 2$ 
as well as $n_{k,j} \geq 0$.

This model
has a dual description \cite{KV} in terms of a certain compactification of the 
heterotic $E_8 \times E_8$ string on first a torus $T_2$ and then 
on $K3$.  This is the so-called heterotic $S$-$T$ model
with
\begin{eqnarray}
S &=& -i  z^1, \hspace{2cm} T \ = \ -i z^2.
\end{eqnarray}
The dilaton $S$ is related to the tree-level coupling constant and to the
theta angle by $S=4 \pi/g^2 - i \theta/2 \pi$.

In order to relate the type II description to its dual heterotic 
description, the 
type II coordinates $t_1$ and $t_2$ must be mapped to the heterotic
coordinates $S$ and $T$.  Based on the physical requirement that
the non-perturbative duality transformations should preserve the
positivity of ${\rm Re} \; S$, it has been argued in \cite{KLM,KKLMV}
that the correct identification is given by 
\beqa
t_1 = S - T \;\;\;,\;\;\; t_2 = T \;\;\;.
\eeqa
In the following, we will take this to be the correct identification.
Thus, 
in the chamber ${\rm Re} \;S > {\rm Re} \;T$, 
the heterotic prepotential\footnote{In the following we will only specify
the prepotential in this particular chamber.} 
reads for $T>1$
\begin{equation}
{\cal F}_{\rm het} = - ST^2 - \alpha T^3 - c 
+ \frac{1}{8 \pi^3} \sum_{j \geq 0, k \geq 1} n_{k,j}
Li_3(e^{-2\pi(jS +(k-j)T)}) 
 - \frac{\beta}{4 \pi^3} Li_3(e^{-2\pi(S-T)}), 
% \nonumber\\
\label{hetprepot}
\end{equation}
where $\alpha = -\frac{1}{3}$ and $\beta 
= - \frac{1}{2}  n_{0,1} = -1$.  The $S$-$T$ model possesses an
$S \leftrightarrow T$ exchange symmetry \cite{KLM}, which is reflected in the 
instanton coefficients which satisfy $n_{k,j}=n_{k,k-j}$ \cite{HoKle}.

At $S = T$, there is a genuine gauge symmetry enhancement \cite{KleMay}.
A $U(1)$ group gets enhanced to an $SU(2)$ 
and four additional hypermultiplets become massless at this point 
in moduli space.  
Three of them belong to the adjoint representation
of $SU(2)$.  The $SU(2)$ can then be completely higgsed away.  On the
type II side this amounts to an extremal transition to a Calabi--Yau
threefold with Hodge numbers $(h^{1,1},h^{2,1})= (1,129)$
and Euler number ${\tilde \chi} = - 256$ \cite{KleMay}.

In the standard perturbative regime $S \rightarrow \infty$ with $T$ finite,
the heterotic prepotential is given by
\beqa
{\cal F}_{\rm het} &=& 
%- ST^2 - \alpha T^3 - c 
%+ \frac{1}{8 \pi^3} \sum_{k \geq 1} n_{k,0}
%Li_3(e^{-2\pi kT}) \nonumber\\
% &=& 
- ST^2 - \alpha T^3 - c 
- \frac{1}{4 \pi^3} \sum_{k \geq 1} c(k)
Li_3(e^{-2\pi kT}) \;\;\;,
\label{standprep}
\eeqa
whith $n_{k,0}= - 2 c(k) > 0 $.  Note that at $T \approx 1$, $\partial^2_T
{\cal F}_{\rm het}$ develops a singularity proportional to $ \log (T -1)$
\cite{KLT}.
In the vicinity of the wall $S=T \rightarrow \infty$, on the other hand,
it follows from
\beqa
Li_3(e^{-x}) &=& p(x) + q(x) \log x \;\;\;,\;\;\; x \rightarrow 0 \;\;\;,
\nonumber\\
p(x) &=& \zeta (3) - \frac{\pi^2}{6} x +\frac{3}{4} x^2 + {\cal O}(x^3) \;\;\;,
\nonumber\\
q(x) &=& - \frac{1}{2} x^2 + {\cal O}(x^3)
\eeqa
that 
\beqa
{\cal F}_{\rm het} &=& - ST^2 - \alpha T^3 - {\tilde c} 
+ \frac{\beta}{2 \pi}(S-T)^2 \log (S-T) \;\;\;,
\eeqa
where ${\tilde c} = c + \frac{\beta \zeta(3)}{4 \pi^3} =
\frac{ {\tilde \chi} \zeta(3)}{16 \pi^3}$.  Here, we have
also omitted terms which are linear and quadratic in $(S-T)$.

Finally, consider writing (\ref{hetprepot}) as
\beqa
 {\cal F}_{\rm het}  &=& -ST^2 \ + \ f(S,T).
\eeqa
Here $f(S,T)$ encodes all perturbative and non-perturbative
quantum corrections and may be expanded in powers
of $e^{-2\pi S}$, as follows 
\cite{CeFe,HenMoo,Hen}
\beqa
 f(S,T)  &=& \sum_{k=0}^{\infty}  f_{k}(T) \ e^{-2\pi k S}, 
\eeqa
where $f_{0}(T)\equiv h(T)$ encodes all the perturbative quantum
corrections in the standard weak coupling limit 
$S \rightarrow \infty$.
It follows 
that  the prepotential $F(Y) = -i (Y^0)^2 
{\cal F}_{\rm het}$ and its periods $F_I(Y)$ are given by 
\beqa
 F(Y)  &=& -i (Y^{0})^{2} 
\left [
-ST^2 \ + \ f(S,T) 
\right ] \;\;\;, \nonumber\\
 F_{0}  &=& i Y^{0}  
\left [
-ST^2  - 2f + T f_{T}  + S f_{S} 
\right ] \;\;\;,
\nonumber\\
 F_{1}  &=& Y^{0}  
\left [ T^2  -  f_{S} \right ] \;\;\;,
\nonumber\\
 F_{2}  &=& Y^{0}  
\left [ 2ST  -  f_{T}  \right ] \;\;\;.
\eeqa
In special coordinates, 
the associated K\"ahler potential reads
\begin{eqnarray}
K(S,{\bar S},T,{\bar T}) &=& 
- \mbox{log} ( S + \bar S + \Delta )
- \mbox{log} ( T + \bar T )^2.
\end{eqnarray}

Here, $\Delta$ contains perturbative and non-perturbative corrections
and is defined as follows:
\begin{eqnarray}
\Delta(S,{\bar S},T,{\bar T}) &=&
\frac{ 2( f + \bar f ) - (T + \bar T)(f_{T} + \bar f_{\bar T})  
 -  (S + \bar S)(f_{S} + \bar f_{\bar S}) }
{ (T + \bar T)^2} \;\;\;.
\end{eqnarray}
In the standard weak coupling limit these corrections reduce to the
Green--Schwarz term \cite{DKLL} 
\begin{eqnarray}
\lim_{S \rightarrow \infty}
\Delta(S,\bar S,T,{\bar T}) &=&
V_{GS}(T,{\bar T}) \ = \
\frac{ 2( h + \bar h ) - (T + \bar T)(h_{T} + \bar h_{\bar T}) }
{ (T + \bar T)^2},
\end{eqnarray}
and in the classical limit these quantum corrections vanish (by definition).
The true target--space duality invariant 
perturbative string coupling constant is given by \cite{DKLL}
\beqa
{8 \pi\over g^2_{pert}}= S + \bar S+V_{GS}(T , \bar T) 
\;.\label{oneloopc}
\eeqa

%%%%%%%%%%%%%%%%%%%%%%%%%%%%%%%%%%%%%%%%%%%%%%%%%%%%%%%%%%%%%%%%%

\subsection{Perturbative target--space duality transformations}

The target--space duality group $\Gamma$ is a certain subgroup of
$Sp(6,{\bf Z})$.  At the perturbative level,
these duality transformations amount to 
 $PSL(2, {\bf Z})_T$ transformations of the modulus $T$, 
which are generated by $ T \rightarrow T + i $ and $T \rightarrow 1/T$.
The latter transformation will be 
of special interest in the following.

Consider the perturbative BPS mass 
\beqa
M^2_{BPS} = |Z|^2 = e^K |q_I X^I - p^I F_I|^2 
=  e^K |M_I P^I + N^I i Q_I|^2 \;\;\;,
\eeqa
where the section $V = (P^I, i Q_J)^T$  is given by
\beqa
V &=& (P^I, i Q_J)^T 
=(1,T^2,iT,i(ST^2 + 2 h - T h_T), iS, -2ST + h_T)^T \;\;\;, \nonumber\\
h&=& 
- \alpha T^3 - c 
- \frac{1}{4 \pi^3} \sum_{k \geq 1} c(k)
Li_3(e^{-2\pi kT}) \;\;\;,
\label{sectionv}
\eeqa
and where 
\beqa
M_I &=& (q_0,-p^1,q_2) \;\;\;, \nonumber\\
N^I &=& (p^0, q_1, p^2) \;\;\;.
\eeqa
The duality transformation $T \rightarrow 1/T$ acts as follows
\cite{KKLMV,HenMoo} on the section $V$ given in (\ref{sectionv})
\beqa
V &\rightarrow& {\bf S}_1 \; V \;\;\;,\;\;\;
 {\bf S}_1 =
\left( \begin{array}{cc}
U & Z\\
W & V\\
\end{array} \right)\;\;\;,\;\;\;
U = \left( \begin{array}{ccc}
0&1&0\\
1&0&0\\
0&0&1\\
\end{array} \right) , \nonumber\\
Z &=& 0 \;\;,\;\;
W=
\left( \begin{array}{ccc}
1&-1&0\\
-1&1&0\\
0&0&0\\
\end{array} \right) \;\;,\;\;
V=
\left( \begin{array}{ccc}
0&1&0\\
1&0&0\\
0&0&1\\
\end{array} \right) \;\;\;.
\label{tdual}
\eeqa
It follows from (\ref{tdual}) that \cite{KKLMV,HenMoo}
\beqa
S &\rightarrow& S - i + \frac{1}{T^2} (2 h - T h_T + i) \;\;\;, \nonumber\\
h & \rightarrow & \frac{h}{T^4}  +  \frac{i}{2T^4}
- \frac{i}{T^2} + \frac{i}{2} \;\;\;, \nonumber\\
h_T & \rightarrow & - \frac{h_T}{T^2} + \frac{4h}{T^3} + \frac{2i}{T^3}
-\frac{2i}{T} \;\;\;.
\label{shht}
\eeqa
Next, consider taking $T$ to be real.  In the region 
${\rm Re} \; T > 1$, 
both $h$ and $h_T$ are real.  Then, it follows from (\ref{shht}) that
\beqa
{\rm Re} \; S &\rightarrow & {\rm Re} \; S +  
\frac{1}{T^2} (2 h - T h_T ) \;\;\;, \nonumber\\
{\rm Re} \; h & \rightarrow & \frac{ {\rm Re} \; h}{T^4} \;\;\;, \nonumber\\
{\rm Re} \; h_T & \rightarrow & - \frac{{\rm Re} \; h_T}{T^2} 
+ \frac{4 {\rm Re} \; h}{T^3} \;\;\;, \nonumber\\
\frac{{\rm Re} \; ( h - T h_T)}{T^2} & \rightarrow &
\frac{{\rm Re} \; ( h - T h_T)}{T^2} \; - \;
 \frac{2 {\rm Re}\; (2h - Th_T)}{T^2} \;\;\;.
\label{rshht}
 \eeqa
In the region 
$ {\rm Re} \; T < 1$, 
on the other hand,
both  $h$ and $h_T$ acquire
imaginary parts, as can be seen from (\ref{shht}).

The charges $(M_I, N^J)$ transform as follows under (\ref{tdual})
\beqa
M \rightarrow U^{T,-1} M - W N \;\;,\;\; N \rightarrow U N \;\;\;.
\label{mntransf}
\eeqa
This should be contrasted with the classical transformation
law, which follows from (\ref{mntransf}) by setting $W=0$.

It will turn out to be convenient to introduce the 
$O(2,1)$ scalar product \cite{CLM}
\beqa
\langle N,N \rangle &=& (N^2)^2 + N^0 N^1  = (p^2)^2 + p^0 q_1 
%\nonumber\\
%\langle 
%M,M \rangle &=& \frac{1}{4} M_2^2 + M_0 M_1 =\frac{1}{4} (q_2)^2 - q_0 p^1
\eeqa
Note that $\langle N,N \rangle$ is invariant under both classical
and perturbative target space duality transformations \cite{CLM}.

The perturbative entropy of $N=2$ supersymmetric 
quantum black holes in the BPS limit is, in 
target--space duality invariant form, given as follows \cite{BCDKLM}
\beqa
\label{per_ent}
S_{BH} &=& \frac{8\pi^2}{g^{2}_{pert}}_{|{\rm fix}} \ \langle N,N \rangle,
\eeqa
with $g_{pert}$ defined in (\ref{oneloopc}) 
and with the fields taking their fixed values on the horizon.

%%%%%%%%%%%%%%%%%%%%%%%%%%%%%%%%%%%%%%%%%%%%%%%%%%%%

\section{Axion-free black holes in the $S$-$T$ model}

\setcounter{equation}{0}

In this section, we will compute the entropy for certain classes
of BPS black hole solutions. We will take $T$ to be real
in the following.
Moreover,  we will first consider 
the region ${\rm Re} \; S > {\rm Re} \; T$ with ${\rm Re} \; T >1$.
Then, it is possible to have perturbative 
axion-free solutions in this region of 
moduli space.  Axion-free solutions are solutions with ${\rm Re} \; z^A = 0$,
that is, ${\rm Im } \; S = {\rm Im } \; T = 0$.  
In the region ${\rm Re} \; T <1$, on the other hand, it is not any longer
possible to set ${\rm Im } \; S = 0$, as can be seen from (\ref{shht}).

For the axion-free solutions in the region 
${\rm Re} \; T > 1$, 
the
extremisation conditions (\ref{extremcon}) yield (with $z^A = Y^A/Y^0$)
% \cite{??}
\beqa
Y^0 = \frac{1}{2}(\lambda + i p^0) \;\;\;,\;\;\; 
z^A \lambda  = i p^A \;\;\;,\;\;\; F_I - {\bar F}_I = i q_I \;\;\;,
\label{cextr}
\eeqa
where $\lambda = Y^0 + {\bar Y}^0$. Thus one can consider three different
cases:
({i}) $\lambda \neq 0, p^0 \neq 0$, 
({ii}) $\lambda =    0, p^0 \neq 0$ and 
({iii}) $\lambda \neq 0, p^0 =    0$.
We will discuss each of these cases in the following.

\subsection{The axion-free $S$-$T$ 
black hole with $\lambda \neq 0, p^0 \neq 0$}

In this subsection, 
we will be interested in perturbative axion-free black hole
solutions in the region $S \gg T>1$ with 
$\lambda \neq 0, p^0 \neq 0$.  The extremisation conditions
(\ref{cextr}) then yield
\beqa
\frac{1}{\lambda^2} &=& \frac{q_1}{p^0 (p^2)^2}, \hspace{1cm}
S \ = \ \frac{p^1}{\lambda} = \frac{p^1}{p^{2}} \sqrt{\frac{q_1}{p^0}}, 
\hspace{1cm}
T \ = \ \frac{p^2}{\lambda} = \sqrt{\frac{q_1}{p^0}}
\eeqa
as well as 
\beqa
q_0 &=& 
- \frac{p^1q_1}{p^0} -  2 \lambda h + \lambda T h_T , \hspace{2cm}
q_2 \ = \  2 \ \frac{p^1q_1}{p^2} - p^0 h_T \;\;\;,
\label{qd}
\eeqa
with $h$ given in (\ref{sectionv}).
For real $T$, $h$ and $h_T$ are also real.
Solving (\ref{qd}) for $h$ and $h_T$ yields
\beqa
h_T &=& 
% {\rm Re} \; h_T = 
2\frac{p^1 q_1}{p^0 p^2} - \frac{q_2}{p^0} \;\;\;, \nonumber\\
h &=&
% {\rm Re } \; h = 
\frac{1}{2 \lambda p^0} \left( p^1 q_1 - p^2 q_2 - p^0 q_0 \right) \;\;\;,
\nonumber\\
2h - T h_T &=& 
% {\rm Re} \; (2h - T h_T) = 
- \frac{1}{\lambda p^0} \left(p^1 q_1 + p^0 q_0 \right) \;\;\;,
\nonumber\\
h - T h_T &=&
% {\rm Re } \; (h - T h_T) = 
\frac{1}{2 \lambda p^0} \left( - 3 p^1 q_1 + p^2 q_2 - p^0 q_0
\right) \;\;\;.
\label{hht}
\eeqa
Note that (\ref{hht}) relates infinite sums over polylogarithmic functions
(appearing on the left hand side)
to simple expressions on the right hand side. Moreover, it is possible to
determine the parameter $\lambda$ completely in terms of the charges, because
the perturbative quantum corrections are independent of the dilaton
($\frac{\partial}{\partial S} h(T) = 0$).
Including non-perturbative corrections encoded in $f(S,T)$ destroys this
property of the black hole solution. In this more general case $\lambda$
remains an undetermined parameter.

Let us now check the target--space duality transformation properties 
of (\ref{hht}) under  $p^0 \leftrightarrow q_1$, that
is, under $T \rightarrow 1/T$.  It follows from (\ref{rshht})
that the left hand side of (\ref{hht}) transforms as  
\beqa
{\rm Re} \; h & \rightarrow & 
% \frac{ {\rm Re} \; h}{T^4} = 
\frac{p^0}{2 q_1^2 \lambda} ( p^1 q_1 - p^2 q_2 - p^0 q_0) \;\;\;,
\nonumber\\
{\rm Re} \; h_T & \rightarrow & 
% - \frac{{\rm Re} \; h_T}{T^2} + \frac{4 {\rm Re} \; h}{T^3} = 
- \frac{q_2}{q_1} - 2 \frac{p^0 q_0}{q_1 p^2} \;\;\;.
\eeqa
The right hand side of (\ref{hht}) transforms
in the same way, provided
% one takes the charges $q$ and $p$ to
the electric and magnetic charges transform as follows:
\beqa
p^0 \leftrightarrow  q_1, \;\;\;\; p^2 \rightarrow p^2, \;\;\;
q_0 \leftrightarrow  -p^1, \;\;\;\; q_2 \rightarrow q_2.
\label{pqcl}
\eeqa
Note that these are the classical transformation laws associated
to $T \rightarrow 1/T$ (cf. eq. (\ref{mntransf})). 
Similarly, it follows from (\ref{rshht}) that $S$ transforms as
\beqa
{\rm Re} \; S &\rightarrow&  
%{\rm Re} \; S +  
%\frac{1}{T^2} (2 h - T h_T) 
%
%=
% \frac{p^1}{\lambda}
%- \frac{\lambda}{p^0 (p^2)^2}( p^1 q_1 +  p^0q_0) =
% 
%\ = \ 
-  \frac{q_{0}}{p^{2}} \sqrt{\frac{p^0}{q_1}} \;\;\;,
\eeqa
which is also consistent with the transformation behaviour of 
$S = p^1/\lambda$ under (\ref{pqcl}).
Note that in the classical limit the dilaton is only invariant 
under target--space duality transformations if 
the additional charge constraints,
given by (\ref{qd}), are taken into account.

The perturbative entropy is then given by (\ref{per_ent}) with
\beqa
\label{g_1}
\frac{8\pi}{g^{2}_{pert}}_{|{\rm fix}} &=& \frac{1}{2} \sqrt{\frac{q_1}{p^0}}
\left (
 \frac{p^1}{p^2} + \frac{q_2}{q_1} - \frac{p^0 q_0}{q_1 p^2}
\right ) \;\;\;.
\eeqa
In the classical limit, on the other hand, we find, for
the dilaton on the horizon, that
\beqa
\frac{4\pi}{g^{2}}_{|{\rm fix}} &=&  \frac{p^1}{p^2} \sqrt{\frac{q_1}{p^0}}
\eeqa
as well as 
the classical duality invariant charge constraints $p^1 q_1=-p^0 q_0 
= \frac{1}{2} p^2 q_2$, which follow from (\ref{qd}).

Note that (\ref{g_1}) was computed in the region ${\rm Re} \; S > 
{\rm Re} \; T > 1$.
It is easy to check that the perturbative
string coupling constant (\ref{g_1}), given in terms of the bare charges on
the horizon, is invariant under the classical
target--space duality transformations (\ref{pqcl}) of the charges.
Thus, the entropy formula 
\beqa  \label{410}
{S}_{BH} = 
\frac{\pi}{2} \sqrt{\frac{q_1}{p^0}}
\left (
 \frac{p^1}{p^2} + \frac{q_2}{q_1} - \frac{p^0 q_0}{q_1 p^2}
\right ) \left(
 (p^2)^2 + p^0 q_1 \right)
\eeqa
actually holds in the entire chamber
${\rm Re} \; S \gg {\rm Re} \; T$.
Note that the entropy varies smoothly across the point 
$T = 1$, where $p^0 = q_1$.

%%%%%%%%%%%%%%%%%%%%%%%%%%%%%%%%%%%%%%%%%%%%%%%%

\subsection{The axion-free S-T black hole with $\lambda = 0, p^0 \neq 0$}

Here, we will be interested in perturbative axion-free black hole
solutions in the region $S \gg T>1$ with 
$\lambda = 0, p^0 \neq 0$.  The extremisation conditions
(\ref{cextr}) then yield $p^A=0, q_0=0$ and
\beqa
\langle N,N \rangle = p^0 q_1 \;, \hspace{1cm}
T = \sqrt{\frac{q_1}{p^0}}\;, \hspace{1cm}
S \ = \ 
% \frac{1}{2} \frac{q_2}{\sqrt{p^0 q_1}} + \frac{h_T}{2T} =
  \frac{1}{2} \frac{q_2}{\sqrt{p^0 q_1}} + \frac{1}{2} \sqrt{\frac{p^0}{q_1}}
\;h_T \;,
\label{st}
\eeqa
with $h$ given in (\ref{sectionv}). 
Under $T \rightarrow 1/T$ we have again 
$p^0 \leftrightarrow  q^1$ and $q_2 \rightarrow q_2$ as in (\ref{pqcl}).
Moreover (\ref{rshht}) also holds for this solution.

The perturbative string coupling constant on the horizon is now given by
\beqa
\label{g_2}
\frac{8\pi}{g^{2}_{pert}}_{|{\rm fix}} &=& 
\frac{p^0}{q_1} 
\left (
 \sqrt{\frac{q_1}{p^0}} \frac{q_2}{p^0} + \mbox{Re} \ h(\sqrt{q_1/p^0})
\right ) \;\;.
\eeqa
It is easy to check that the perturbative
string coupling constant (\ref{g_2}) is indeed invariant under the classical
target--space duality transformations (\ref{pqcl}) of the charges.
It follows that the perturbative entropy formula 
\beqa
{S}_{BH} = \pi \; \left(
\sqrt{p^0 q_1 q_2^2} + (p^0)^2 {\rm Re} \; h(\sqrt{q_1/p^0}) \right)
\label{perbh}
\eeqa
holds in the entire chamber ${\rm Re} \; S \gg {\rm Re} \; T$.
Note again that the entropy (\ref{perbh}) varies smoothly across the point 
$T = 1$, where $p^0 = q_1$. 

In the classical limit 
% the classical constraint $p^1 q_1 = - p^0 q_0$ holds again and 
the string coupling constant on the horizon and 
the classical entropy have the following form:
\beqa
\frac{8\pi}{g^{2}}_{|{\rm fix}} = 
\sqrt{\frac{p^0}{q_1}} \frac{q_2}{p^0} \;\;\;, \;\;\;
S_{BH}^{class} &=& \pi \ \sqrt{p^0 q_1 q_2^2} \;\;\;.
\eeqa

%%%%%%%%%%%%%%%%%%%%%%%%%%%%%%%%%%%%%%%%%%%%%%%%%%%%%%%%%%%%

\subsubsection{The entropy in the limit $ S \approx T \rightarrow 0$}

In the strong coupling limit $ S \approx T \rightarrow 0$ the heterotic
prepotential is given by $f(S,T)$ only. In particular
we find
\beqa
{\cal F}_{\rm het} &=& f(0,0) \ = \
\frac{1}{8 \pi^3} \zeta(3)\sum_{j,k \geq 0} n_{k,j} 
\label{prepst0}
\eeqa
with $n_{0,0}= -8 \pi^3 c - 2 \beta$. Since the sum is divergent, 
this expression is only to be understood in an asymptotic sense. For vanishing
$S$ and $T$ the prepotential would diverge, because of this infinite
sum. In this limit, the entropy is then given by
\beqa
S_{BH} = \pi \; \left( |Y^0|^2 e^{-K} \right)_{\rm | fix} =
(p^0)^2 \frac{1}{8 \pi^2} \zeta(3)\sum_{j,k \geq 0} n_{k,j}  \;\;\;.
\label{entzep}
\eeqa In section \ref{micr}, we will discuss a microscopic
interpretation for the entropy (\ref{entzep}).

%%%%%%%%%%%%%%%%%%%%%%%%%%%%%%%%%%%%%%%%%%%%%%%%%%%%%%%%%%%%%

\subsection{The axion-free $S$-$T$ black hole with $\lambda \neq 0, p^0 = 0$}

Next, we
will be interested in perturbative axion-free black hole
solutions in the region $S >T$ with 
$\lambda \neq 0, p^0 = 0$.  More precisely, we will discuss the
general standard weak coupling limit, the general black hole 
solution including non-perturbative corrections and a special
weak coupling limit near $S=T$. 
%////// I WOULD REMOVE THIS:
%While the first two cases
%can only be solved implicitly, the last one can be solved explicitly,
%In the latter case
%we will be able to solve explicitly for the entropy.
% provided we take the
%fixed points of $S$ and $T$ to lay in the vicinity of the wall
%$S = T$.  
%since 
%///////////////
This case is analogous to one studied in the context of the
$S$-$T$-$U$ model, where the fixed points of $T$ and $U$ had to be
taken to lay near the wall $T = U$ of perturbative gauge symmetry
enhancement \cite{be/ga}.  In the case of the $S$-$T$ model, there is
a genuine gauge symmetry enhancement on the wall $S=T$ \cite{KleMay}.
We will use the $S \leftrightarrow T$ exchange symmetry of the model
in order to determine the entropy in the two chambers ${\rm Re} \; S >
{\rm Re} \; T$ and ${\rm Re} \; S < {\rm Re} \; T$ near the wall
$S=T$.

Recall that, in the chamber ${\rm Re} \; S > {\rm Re} \; T$,
the heterotic prepotential is given by 
${\cal F}_{\rm het} = - ST^2 + f(S,T)$ with
\beqa
f(S,T) &=& - \alpha T^3 - c 
+ \frac{1}{8 \pi^3} \sum_{j \geq 0, k \geq 1} n_{k,j}
Li_3(e^{-2\pi(jS +(k-j)T)}) -  \frac{\beta}{4 \pi^3}
Li_3(e^{-2\pi(S-T)}) .
\nonumber\\ 
\label{fhetst}
\eeqa
For the case $\lambda \neq 0, p^0 = 0$, 
the extremisation conditions
(\ref{cextr}) then yield
\beqa
q_A &= & 0 
\;\;\;,\;\;\; Y^A = i \frac{p^A}{2} \;\;\;,\;\;\; 
Y^0 = \frac{\lambda}{2} \;\;\;,\;\;\; z^A = i \frac{p^A}{\lambda}\;\;\;.
\eeqa
The parameter $\lambda$ is determined (in general implicitly) by the constraint
\beqa
i q_0 &=& 4 \frac{\partial}{\partial \lambda} F(\lambda, p^A)\;\;\;.
\label{constrlam}
\eeqa

%%%%%%%%%%%%%%%%%%%%%%%%%%%%%%%%%%%%%%%%%%%%%%%%%%%%%%%%%%%%%%%%

\subsubsection{Standard weak coupling limit}

In the standard weak coupling limit $S=p^1/\lambda \rightarrow \infty$
with arbitrary but finite $T=p^2/\lambda$, and consequently
$f(S,T)\rightarrow h(T)$, the entropy is given by (\ref{per_ent})
with 
\beqa
\label{g_3}
\frac{8\pi}{g^{2}_{pert}}_{|{\rm fix}} &=& 
2 \frac{p^1}{\lambda}
+ \frac{\lambda^2 }{ (p^2)^2} h(p^2/\lambda)
-  \frac{\lambda}{p^2} h_{T}(p^2/\lambda) \;\;\;.
\eeqa
Using (\ref{rshht}) it is easy to show that the perturbative
string coupling constant on the horizon (\ref{g_3}) is invariant
under target--space duality transformations
$T \rightarrow 1/T$ with
\beqa
\label{charge_3}
\lambda \rightarrow (p^2)^2/\lambda, \hspace{1cm}
p^2 \rightarrow p^2, \hspace{1cm}
\frac{p^1}{\lambda}  \rightarrow 
 \frac{p^1}{\lambda} + 
 \frac{\lambda^2}{(p^2)^2} (2h - \frac{p^2}{\lambda} h_{T}).
\eeqa
In the classical limit these transformations reduce
to $p^1 \leftrightarrow q_{0}$ and $p^2 \rightarrow p^2$.
The string coupling constant has the
following fixed value in terms of the charges on the horizon:
\beqa
\frac{4\pi}{g^{2}}_{|{\rm fix}} &=& \frac{p^1}{\lambda}, \hspace{2cm}
\lambda \ = \ \sqrt{-\frac{p^1 (p^2)^2}{q_0}}.
\eeqa
Thus, the classical entropy of the black hole in terms of the 
charges is ($q_0<0$)
\beqa
\label{ent_2_class}
S_{BH}^{class} &=& \pi \ \sqrt{|q_0| p^1 (p^2)^2}.
\eeqa

%%%%%%%%%%%%%%%%%%%%%%%%%%%%%%%%%%%%%%%%%%%%%%%%%%%%%%%%%%%%%%

\subsubsection{More general axion-free quantum black holes}

Let us now consider more general quantum corrected black hole solutions
given in terms of $f(S,T)$ and of 
\beqa
f_{S}(S,T) &=& - \frac{1}{4 \pi^2} \sum_{j \geq 0, k \geq 1} n_{k,j} j
Li_2(e^{-2\pi(jS +(k-j)T)}) +  \frac{\beta}{2 \pi^2}
Li_2(e^{-2\pi(S-T)}) \;\;\;,
\nonumber\\
f_{T}(S,T) &=& - 3\alpha T^2
- \frac{1}{4 \pi^2} \sum_{j \geq 0, k \geq 1} n_{k,j} (k-j)
Li_2(e^{-2\pi(jS +(k-j)T)}) -  \frac{\beta}{2 \pi^2}
Li_2(e^{-2\pi(S-T)}). 
\nonumber\\
\eeqa
The corresponding general axion-free black hole entropy 
in the $S$-$T$ model is then given by
\beqa  \label{425}
S_{BH} &=& \pi \; \left(|Y^0|^2 e^{-K}\right)_{| {\rm fix}} = 4 \pi |Y^{0}|^2 
\left ( 2ST^2 + f - T f_T -S f_S \right )_{| {\rm fix}} \;\;\;.
\eeqa
This entropy contains all perturbative and non-perturbative 
quantum corrections encoded in $f(S,T)$ and represents the
general axion-free entropy in the $S$-$T$ model.
For the case considered here we have 
$Y^0=\lambda/2$, $S=p^1/\lambda$ and $T=p^2/\lambda$, 
and the parameter $\lambda$ is subject to the constraint
\beqa
\label{c_1}
q_0 &=& - \frac{p^1(p^2)^2}{\lambda^2} -2 \lambda f - \lambda^2
        \frac{\partial}{\partial \lambda} f
\eeqa
with
\beqa
\frac{\partial}{\partial \lambda} f &=& 
3\alpha \frac{(p^2)^3}{\lambda^4} -  \frac{\beta}{2 \pi^2}
\frac{p^1-p^2}{\lambda^2}
Li_2(e^{-2\pi(p^1-p^2)/\lambda})
\nonumber\\ & & 
+ \frac{1}{4 \pi^2 \lambda^2} \sum_{j \geq 0, k \geq 1} n_{k,j} 
\left (
j p^1 + (k-j) p^2
\right )
Li_2(e^{-2\pi(jp^1 +(k-j)p^2)/\lambda}). 
\eeqa
The constraint (\ref{c_1}) can be solved for a special weak
coupling limit as we will show next.

%%%%%%%%%%%%%%%%%%%%%%%%%%%%%%%%%%%%%%%%%%%%%%%%%%%%%%%%%%%

\subsubsection{Special weak coupling limit}

For the present case $Y^0- \bar Y^0=0$, using (\ref{bh_ent}), 
the general axion-free
entropy can be brought into the following form:
\beqa
S_{BH} &=& \frac{\pi}{2} \Big( - \lambda q_0
+3 \left(\frac{p^1 (p^2)^2}{\lambda} + \alpha \frac{(p^2)^3}{\lambda}
\right) - \frac{\beta \lambda}{2 \pi^2} (p^1-p^2) 
Li_2(e^{- 2 \pi(p^1-p^2)/\lambda}) 
\nonumber\\
& & + \frac{\lambda}{4 \pi^2} 
\sum_{j\geq 0,k \geq 1}
n_{k,j} \left ( (k-j) p^2 + j p^1 \right ) 
Li_2(e^{- 2 \pi((k-j)p^2 + jp^1)/\lambda}) \Big) \;\;\;.
\label{bhentro}
\eeqa

In the perturbative regime 
$ S > T \rightarrow \infty, S - T \approx 0$, that is, 
in the vicinity of the wall
$S=T$ the constraint (\ref{c_1}) can be solved approximately:
\beqa
\label{cc_1}
q_0 \lambda^2 &=& - p^1(p^2)^2 - \alpha (p^2)^3 
+ \lambda^3 \left ( 2c + \frac{\beta \zeta (3)}{2 \pi^3} \right )
- \frac{\beta}{12} (p^1-p^2) \lambda^2 
- \frac{\beta}{2\pi} (p^1-p^2)^2 \lambda \nonumber\\
&+& \cdots \;\;\;.
\eeqa
Here we expanded in $x=2\pi(p^1-p^2)/\lambda$ around $x=0$ using
\beqa
Li_3(e^{-x}) &=& \zeta (3) - \frac{\pi^2}{6} x +  
                 \left(\frac{3}{4} - \frac{\log x}{2} \right) x^2 
                 +  {\cal O}(x^3), \nonumber\\
Li_2(e^{-x}) &=& \frac{\pi^2}{6} +  
                 \left( \log x - 1 \right) x + \frac{1}{2} x^2 
                 +  {\cal O}(x^3).
\eeqa
Note that the logarithmic contributions from the polylogarithms
cancel against each other in (\ref{cc_1}). 
Using that $|p^1-p^2| \ll |p^A| \ll |q_0|$, one can solve
(\ref{cc_1}) in terms of the following
power series expansion
\beqa
\lambda = \sum_{i=1}^{\infty} \frac{\gamma_i}{(\sqrt{q_0})^i}
= \frac{\gamma_1}{\sqrt{q_0}} + \frac{\gamma_2}{q_0} + \dots \;\;\;.
\label{power}
\eeqa
Inserting (\ref{power}) into (\ref{cc_1}) and comparing terms
yields 
\beqa
\gamma_1^2 &=& - p^1 (p^2)^2 - \alpha (p^2)^3, \hspace{2cm}
\gamma_2  \ = \ \frac{\beta}{4 \pi} (p^1-p^2)^2 \;\;\;.
\eeqa
Choosing again $q_0 < 0, p^A > 0$ it follows that
\beqa
\lambda = \sqrt{\frac{- p^1 (p^2)^2 - \alpha (p^2)^3}{q_0}}
+ \frac{\beta}{4 \pi} \frac{(p^1-p^2)^2}{q_0} + \cdots \;\;\;.
\label{exlam}
\eeqa
In the limit  $S > T \rightarrow \infty, S - T \approx 0$, the only 
polylog term contributing to the corresponding quantum corrected
entropy is the term
$ Li_2(e^{- 2 \pi (S-T)}) \approx 2 \pi (S-T) \log  2 \pi (S-T) $.
It follows that in this limit the entropy can be written as
\beqa
{S}_{BH} = 2 \pi \sqrt{|q_0| \left(p^1 (p^2)^2 + \alpha (p^2)^3 
\right)} - \frac{\beta}{4}(p^1-p^2)^2 
\log  
\left( 
 \frac{|q_0| (p^1-p^2)^2}{\left(p^1 (p^2)^2 
 + \alpha (p^2)^3 \right)}
\right) + \cdots
\label{entrosbt}
\eeqa
Equation (\ref{entrosbt}) gives the entropy in the chamber
${\rm Re } \; S > {\rm Re } T$ near the wall ${\rm Re } S= {\rm Re }T$.
By utilising the $S \leftrightarrow T$
exchange symmetry of the model, 
it follows that in the chamber 
 ${\rm Re } \; T >  {\rm Re } S$
the entropy near the wall ${\rm Re }T  = {\rm Re } S$ is given by
(\ref{entrosbt}) with $p^1\leftrightarrow p^2$.
Note, in particular, that the entropy is finite on the wall $S=T$ and that it
varies continuously across the wall $S=T$.
A similar effect, which we will briefly describe next,
% shortly,
also occurs in $5$ dimensions when considering 
the entropy density of the associated 
black string.  
%%%%%%%%%%%%%%%%%%%%%%%%%%%%%%%%%%%%%
%\subsubsection{ The entropy in the limit $ S \approx T \rightarrow 0$}
%
%In the strong coupling limit $ S \approx T \rightarrow 0$, the heterotic
%prepotential is given by
%\beqa
%{\cal F}_{\rm het} = f(0,0) \ = \
%\frac{1}{8 \pi^3} \zeta(3)\sum_{j,k \geq 0} n_{k,j} 
%\eeqa
%The extremisation conditions then yield
%$q_0 = - 2 \lambda f(0,0)$. Thus, we find 
%\beqa
%\lambda = - \frac{ 4 \pi^3 q_0}{\zeta(3)\sum_{j,k \geq 0} n_{k,j}} 
%\eeqa
%and the entropy is then given by \cite{GaiBe}
%\beqa
%{S}_{BH} &=& \pi |Y^0|^2 e^{-K} = (q^0)^2  \frac{2\pi^4}{
%\zeta(3)\sum_{j,k \geq 0} n_{k,j}} \;\;\;.
%\label{entzeq}
%\eeqa
%In section \ref{micr}, we will propose a microscopic explanation
%for the entropy (\ref{entzeq}) [DO WE HAVE ONE ?].
%%%%%%%%%%%%%%%%%%%%%%%%%%%%%%%%%%%%%%%%%%%%%%%%%%%%%%%

%%%%%%%%%%%%%%%%%%%%%%%%%%%%%%%%%%%%%%%%%%%%%%%%%%%%%%%%%%%%%%%%%%%%%

\subsection{The entropy density for the associated $5$ dimensional
black string}

Consider the $S$-$T$ model with the following prepotential (in the chamber
$S>T$)
\beqa
{\cal F} = - \left( ST^2 + \alpha T^3 \right) \;\;\;,\;\;\; \alpha = - 
\frac{1}{3} \;\;\;.
\eeqa
Let $R_5$ denote the radius of the circle of the compactified $5$-th
dimension.  Then \cite{AnFeTa}
\beqa
S = R_5 s \;\;\;,\;\;\; T = R_5 t \;\;\;,
\eeqa
where $s$ and $t$ denote the two moduli fields in $5$ dimensions.
It follows that
\beqa
{\cal F} &=& - R^3_5\; {\cal V} \;\;\;, \nonumber\\
{\cal V} &=&  d_{\Lambda \Delta \Sigma} t^{\Lambda}
t^{\Delta} t^{\Sigma} =
st^2 +  \alpha t^3 \;\;,\;\; t^{\Lambda} = s, t \;\;\;,
\eeqa
where ${\cal V}$ denotes the prepotential of real special
geometry in $5$ dimensions.  It has to 
satisfy the additional constraint \cite{AnFeTa}
\beqa
{\cal V} = 1 \;\;\;.
\label{veqo}
\eeqa
Consider a black string carrying charges
$p^1$ and $p^2$.
In the $M$-theory picture these charges are magnetic and carried by two
5-branes, whereas in the heterotic picture $p^1$ is the charge of the
fundamental string, i.e.\ electric, and $p^2$ comes from a 5-brane 
that has been identified with a KK-monopole (see next section).
%%%%%%%%%%%%%%%%%%%%%%%%%%%%%%%%%%%%%%%%%
%denote the charge of the fundamental string 
%($p^1$) and
%arise from ten dimensional five-branes wrapping around $K3$ ($p^2$)
%\cite{five_branes}. In the eleven dimensional supergravity theory
%($M$-theory) these charges have their origin in five-brane solitons,
%which wrap even cycles in the Calabi-Yau space
%\cite{5_soliton}
%\beqa
%p^{\Lambda} &=& \int_{C_{2}^{\Lambda} \times S_{2}} \ F_4.
%\label{cycles}
%\eeqa
%Here $F_4$ is the field strength of the antisymmetric tensor in eleven
%dimensional supergravity and $C_{2}^{\Lambda}$ are 4-cycles
%in the Calabi-Yau space. 
%%%%%%%%%%%%%%%%%%%%%%%%%%%%%%%%%%%%%%%%%%%%
The associated magnetic central charge
is given by \cite{AnFeTa}
\beqa
Z_m = - t_{\Lambda} p^{\Lambda} 
= - \left( p^1 t^2 + p^2  (2st + 3 \alpha t^2)\right) \;\;\;,
\label{zm}
\eeqa
with $ t_{\Lambda}= d_{\Lambda \Delta \Sigma} t^{\Delta} t^{\Sigma}$. 
It can be derived from the central charge in $4$ dimensions, as
follows.  The $4$ dimensional central charge reads
\beqa
Z^{4D} = e^{K/2} {\cal M} \;\;\;,\;\;\; {\cal M} = q_I X^I - p^I F_I \;\;\;.
\eeqa
For the case at hand, $p^0 = 0, q_A = 0$, so that
\beqa
{\cal M} =  R_5^2 \left( \frac{q_0}{R_5^2} - 
 p^1 t^2 - p^2  (2st + 3 \alpha t^2) \right) \;\;\;.
\eeqa
The magnetic central charge $Z_m$ in $5$ dimensions is related
to the $4$ dimensional central charge in the following way
\beqa
Z_m &=& \lim_{R_5 \rightarrow \infty} \ (R_5)^{-1/2}  \ Z^{4D} 
\nonumber\\
&=&   \lim_{R_5 \rightarrow \infty}
(R_5)^{3/2}  \ e^{K/2} \left(\frac{q_0}{R_5^2} - 
 p^1 t^2 - p^2  (2st + 3 \alpha t^2) \right).
\eeqa
Using that $K = - \log (2 R_5)^3 - \log {\cal V} = 
- \log (2 R_5)^3 $, it follows that 
\beqa
Z_m &=& \lim_{R_5 \rightarrow \infty} (R_5)^{-1/2} Z^{4D} =
 - p^1 t^2 - p^2  (2st + 3 \alpha t^2) \;\;\;,
\eeqa
up to an overall constant factor.  This is
in accordance with (\ref{zm}).

Inserting the constraint (\ref{veqo}) into (\ref{zm})
yields
\beqa
Z_m = - \left( t^2 (p^1 + \alpha p^2) + \frac{2 p^2}{t} \right) \;\;\;.
\eeqa
According to \cite{FerKal1}, the entropy density can be obtained
by solving the extremization condition
\beqa
\frac{\partial}{\partial t} Z_m = 0 \;\;\;.
\label{extrcond}
\eeqa
The extremisation condition (\ref{extrcond}) yields
\beqa
t^3 = \frac{p^2}{p^1 + \alpha p^2} \;\;\;.
\label{tfix}
\eeqa
Inserting (\ref{tfix}) into (\ref{zm}) yields 
the magnetic central charge
at the fixed point
\beqa
Z_m|_{\rm fix} &=& - 3 \left( (p^2)^2 (p^1 + \alpha p^2) \right)^{1/3}
\ = \ - 3 \left( d_{\Lambda \Delta \Sigma} p^{\Lambda}
p^{\Delta} p^{\Sigma} \right)^{1/3} \;\;\;.
\eeqa

The $5$ dimensional entropy density\footnote{Note that, for 
extended objects, one usually considers densities instead of
total quantities. In analogy to the BPS mass density it is reasonable
to discuss the entropy density here.} is then given by 
%%%%%%%%%%%% KILL %%%%%%%%%%%%%%%%%%%
\beqa \label{f_ent}
{S}_{BH}^{5D} &\propto& Z_m^2|_{\rm fix} \ \propto \ 
 \left( d_{\Lambda \Delta \Sigma} p^{\Lambda}
p^{\Delta} p^{\Sigma} \right)^{2/3} \;\;\;.
\eeqa
This is the entropy density in the chamber $s>t$.  In the chamber $t>s$,
on the other hand, (\ref{f_ent}) holds with $p^1 \leftrightarrow p^2$.  
%%%%%%%%%%%%%%%%%%%%%%%%%%%%%%
%\beqa
%{\cal S}^{5D} \propto Z_m^2|_{\rm fix} \propto
%\left( - p^2 (p^1)^2 - \alpha (p^1)^3  \right)^{2/3} \;\;\;.
%\eeqa
%%%%%%%%%%%%%%%%%%%%%%%%%%%%%%
Hence, it follows that
\beqa
{S}_{BH}^{5D} &\propto&
\left( p^1 (p^2)^2  + \alpha (p^2)^3 \right)^{2/3}\theta(s-t)
+
\left(  p^2 (p^1)^2 + \alpha (p^1)^3  \right)^{2/3} \theta(t-s) \;\;\;.
 \eeqa

%%%%%%%%%%%%%%%%%%%%%%%%%%%%%%%%%%%%%%%%%%%%%%%%%%%%%%%%%%%%%%%%%

\section{The heterotic and type II solutions}

\setcounter{equation}{0}

\noindent
In the previous sections we computed the entropy, by solving a set of
extremisation conditions, for certain classes of black hole solutions.
In this section, we will describe the corresponding black hole and black
string solutions.  On the heterotic side, we have pure Neveu-Schwarz
($NS$) solutions, whereas on the type II side they represent intersections of
$D$- or $M$-branes living in a gas of closed type II strings or closed
$M$-2-branes.

The general $S$-$T$ model allows 6 non-vanishing charges. The
restriction to the axion-free case gives two constraints, given in
equation (\ref{qd}). Thus, in the axion-free case we have only 4
independent charges. In the case (i) we kept all 4 charges. The other
cases (ii) and (iii) are the least charge configuration where we
turned off $q_0$ or $p^0$. These cases are especially interesting.  In
11 dimensions the case (ii) describes an intersection of membranes and (iii)
an intersection of 5-branes.  In 4 dimensions both solutions are
$S$-dual to one another. The general configuration, where we keep the
charges $q_0$ and $p^0$ non-vanishing, describes an interpolation
between these least charged solution. We will in the following focus
on the case (iii), although on the type II side we will present some
speculations about the general solution (case (i)).

We begin with a discussion of the heterotic solutions.
On the heterotic side we can only give a microscopic interpretation to the 
classical solutions. For simplicity we will restrict ourselves to the
special case where $p^0=0$. In this case it follows from 
(\ref{entrosbt}) that the classical entropy is given by
\beqa
\label{020}
{S}_{BH}^{class} = 2 \pi \sqrt{ |q_0 \, p^1 \, p_2^2| } \;\;\; .
\eeqa
Note that, on the heterotic side, $p^1$ is an electric charge.
The corresponding solution in 10 dimensions is given by 
(see the second ref.\ of \cite{D4N4Entr})
\beqa
\label{030}
 \ba{l}
ds_{10}^2 = {1 \over H_1} du \left( dv + H_0 du \right) + dy_m dy_m +
 H_2 \left({1 \over H_2} (dx_8 + \vec{V} d\vec{x})^2 + H_2 d\vec{x} 
 \right) \\
 H =  d(1/H_1) \wedge du \wedge dv + ^*dH_2 \wedge du \wedge dv
  \wedge (dy)_m \qquad , \qquad e^{-2 \hat{\phi}} = {H_1 \over H_2} \;\;\;,
   \ea
\eeqa
($\epsilon_{ijk} \partial_j V_k = \partial_i H_2$, $u,v = x_9 \pm t$, $m=
1..4$). This configuration describes a fundamental string lying in a $NS$
5-brane. In addition, there are momentum modes travelling along the
string (boost), and in the transversal space is a KK-monopole. In
comparison to the $S$-$T$-$U$ model we have identified $T=U$, which means
that the harmonic functions related to the 5-brane and to the KK-monopole
part have been identified. As a consequence, this classical solution is
$T$-selfdual with respect to the $x_8$ direction, but concerning the $u$
direction this duality transformation exchanges $H_0$ with $H_1$ ($q_0
\leftrightarrow p^1$).  When compactifying this solution, one reduces
first over the torus ($x_8 , x_9$). This yields a black hole lying in a
4-brane. In a second step one wraps the 4-brane completely over a $K3$
manifold..  The associated scalar fields are then given by
\beqa
\label{040}
S = e^{-2 \phi} = e^{-2 \hat{\phi}} \sqrt{|G_{rs}|} = \sqrt{H_0 H_1
 \over H_2^2} \quad , \quad T= \sqrt{|G_{rs}|} = \sqrt{H_0 \over H_1} \;\;\;,
\eeqa
where $G_{rs}$ denotes the $(x_8 , u)$ part of the metric (\ref{030}).
For the 4d metric in the Einstein frame one obtains
\beqa
\label{050}
ds^2 = - {1 \over \sqrt{H_0 H_1 H_2^2}} dt^2 + \sqrt{H_0 H_1 H_2^2}
\, d\vec{x} d\vec{x} \ .
\eeqa
Thus, by inserting the harmonic functions $H_0 = 1 + {\sqrt{2}q_0 \over
r}$, $H_1 = 1 + {\sqrt{2} p^1 \over r}$, $H_2= 1 + {\sqrt{2} p^2 \over
r}$ into (\ref{050})
and by calculating the area of the horizon, one obtains the entropy
(\ref{020}). In addition, the scalar fields behave smoothly and take
fixed values on the horizon ($r=0$), which are given in terms of the charges
only.

\smallskip

In addition to the quantum corrections (higher genus corrections),
described in the previous sections, one has to consider $\alpha'$
corrections as well. These terms do not appear in the prepotential,
instead they are related, e.g., to higher curvature corrections.  In
order to have control over these terms as well, we have to make sure that
the curvature in the string frame does not blow up on the horizon. Since
the radius of the horizon in the string frame is proportional to the
magnetic charge, we can suppress these terms by choosing a sufficiently
large charge $p^2$.

\smallskip

% Since the magnetic $NS$ charge does not allow for a flat space
% description [WHAT DOES THAT MEAN?], it is difficult to give the
% Bekenstein-Hawking entropy a statistical interpretation. A possible
% approach has been discussed in \cite{la/wi}. On the other hand, on the
% type II side not only the polynomial part in the prepotential but also
% all the polylogarithmic corrections have a statistical interpretation.

Next, we would like to discuss the solutions on the type II side. The
heterotic solution discussed above can be mapped onto the type II side,
where the corresponding black hole solution can be interpreted as a
compactification of intersecting branes. Both solutions are equivalent,
but on the type II side the corrections to the prepotential have a clear
geometrical interpretation in terms of the Calabi--Yau threefold. Thus,
on the type II side, one can identify the additional states and the
statistical interpretation of the entropy is especially clear.

The black hole becomes non-singular if 4 branes intersect each other. If
one has less branes intersecting each other, the horizon shrinks to zero
size and the black hole becomes singular. Actually, it is not necessary
to have additional branes at the intersection, also internal waves
(boosts) or KK-monopoles can stabilize the horizon.

We will now discuss the type II analogue of (\ref{050}), and we will
mainly do this in the $M$-theory picture. Since the solution has only two
K\"ahler class moduli, we can only wrap two inequivalent branes around
the two non-homologous $4$-cycles of the Calabi--Yau threefold. Since on
the type II side $p^1$ and $p^2$ are magnetic charges, the 11-d brane
configuration must contain the intersection of two 5-branes.  These two
5-branes intersect over a 3-brane and in order to obtain the electric
charge $q_0$, we make again a boost along the worldvolume of the
intersection, i.e.\ along one of the 3-brane directions.  The
corresponding metric is given by (in the following we will mainly
consider the metric) \cite{ts}
\beqa
\label{060}
ds_{11}^2 = {1 \over (H^1 H^2)^{1/3}} \left[ dudv + H_0 du^2 + dx_5 dx_5
 +dx_6 dx_6 +H^1 H^2 d\vec{x} d\vec{x} + H^{\Lambda} \omega_{\Lambda}  
\right]  \;\;\;, 
\eeqa
where $\omega_{\Lambda}$ ($\Lambda=1,2$) are two 2-dimensional
 line elements and 
where $\vec{x}=(x_1
,x_2, x_3)$.  The boost direction is $x_4$ ($u,v = x_4 \pm t$), which is
parametrized by $H_0$. The location of the branes can be chosen as follows:
\renewcommand{\arraystretch}{1}
\beqa
\label{070}
\ba[c]{cccccccccccc}
\hline
&t & x_1 & x_2 & x_3 & x_4 & x_5 & x_6 & x_7 & x_8 & x_9 & x_{10} \\
\hline
boost & \circ &&&& \circ &&&&& \\
H_1 - 5-brane & \times &&&& \times & \times & \times & \times & \times & &\\
H_2 - 5-brane & \times &&&& \times & \times & \times & &&\times & \times \\
\hline
\ea
\eeqa
where the worldvolume coordinates are indicated by ``$\times$'', and
where ``$\circ$'' denotes the boost directions.
\renewcommand{\arraystretch}{1.8}

Next, one has to compactify this configuration on a Calabi--Yau threefold,
which yields a string solution in 5 dimensions ($t , x_1 \ldots x_4$).
Ignoring the instanton corrections for a moment, this solution is given by
\cite{be}
\beqa
\label{072}
ds_{5D}^2 = {1 \over (d_{\Lambda \Delta \Sigma} H^{\Lambda} 
 H^{\Delta} H^{\Sigma})^{1/3}} 
 \left[ dudv + H_0 du^2 + d_{\Lambda\Delta\Sigma} H^{\Lambda} 
 H^{\Delta} H^{\Sigma} \, d\vec{x} d\vec{x}  \right]   \;\;\;,
\eeqa
where $\Lambda, \Delta, \Sigma = 1,2$, and where
$d_{\Lambda\Delta\Sigma}$ denote the intersection numbers of the
Calabi--Yau given in (\ref{preptype2}). The index $\Lambda$ counts the
number of non-trivial 4-cycles of the Calabi--Yau, and in the solution it
indicates around which 4-cycle we have wrapped the 5-brane. For this
string we can define an entropy density (entropy per string length)
and we obtain, after inserting the harmonic functions given after
equation (\ref{050}),
\beqa
\label{073}
{S}_{BH}^{5D} = 2 \pi \, (d_{\Lambda\Delta\Sigma} 
 p^{\Lambda} p^{\Delta} p^{\Sigma})^{2/3} \;\;\;
\eeqa
which coincides with (\ref{f_ent}). 
In a second step one has to compactify this string, that is one has to wrap it
around the $4$-th direction. As a result we obtain the 4-d black hole
\beqa
\label{074}
ds_{4D}^2 = -{1 \over \sqrt{ H_0 \, d_{\Lambda\Delta\Sigma} 
H^{\Lambda} H^{\Delta} H^{\Sigma}}} dt^2 + \sqrt{H_0 \, 
d_{\Lambda\Delta\Sigma} H^{\Lambda} H^{\Delta} H^{\Sigma}} \, 
d\vec{x} d\vec{x}  \;\;\;,
\eeqa
whose 4-d entropy is given by
\beqa
\label{075}
{S}^{4D}_{BH} = 2 \pi \, \sqrt{|q_0| \, d_{\Lambda\Delta\Sigma} p^{\Lambda}
p^{\Delta} p^{\Sigma}} \, \ .
\eeqa
which coincides with the first term in (\ref{entrosbt}).

This model shows that compactifying on a Calabi--Yau threefold can
stabilize a solution. Since the model under consideration has only two
K\"ahler class moduli, we can only wrap two topological inequivalent
(e.g.\ orthogonal) 5-branes around 4-cycles of the Calabi--Yau threefold.
As a consequence all triple intersections are self-intersections, which
stabilize the black hole solution.  Although the 11-d configuration is
singular, the Calabi--Yau compactification makes it non-singular. Because
the dependence of the black hole solution on the intersection numbers is
given via the expression $d_{\Lambda\Delta\Sigma} H^{\Lambda} H^{\Delta}
H^{\Sigma}$, a self-intersection of branes has qualitative the same
consequence as the triple intersection of different branes. Thus, in the
same way as additional branes, also the self-intersections improve the
singularity structure of a black hole.

Finally, we would like to comment on the interpolating case (i).  We
will again only discuss the intersection part of the solution. In this
case, in addition to the magnetic charges we have the electric charges
$q_1, q_2$ as well as the constraints (\ref{qd}). Therefore, we need
in 11 dimensions a brane solution that interpolates between the
2-brane and the 5-brane.  This solution is known and is given by
\cite{iz/la}
\beqa
\label{080}
ds^2 = {1 \over (H \tilde{H})^{2/3}} \left[ \tilde{H} ({\bf M_3} ) + H 
( {\bf E_3}) + \tilde{H} H ({ \bf E_5}) \right] \;\;\;,
\eeqa
where ${\bf M_3}$ and ${\bf E_{n}}$ denote a $3$-dimensional 
Minkowskian and an $n$-dimensional
Euclidian space, respectively. The harmonic functions $H$ and $\tilde{H}$
are function of the transversal space ${\bf E_5}$: $H= 1+ {q \over r^3}$,
$\tilde{H} = 1 + {q \cos^2 \xi \over r^3}$.  For $\xi =0$ we have a
5-brane, and for $\xi = \pi/2$ we have a 2-brane. We do not wish to
discuss this solution in detail, but we would like to point out that an
intersecting configuration in terms of these objects along the line of
\cite{co} could provide a microscopic picture for the case (i).

In order to understand the instanton corrections to this solution, it
is not sufficient to consider the 11-d intersection of 5-branes and
their compactification only. Instead, one has to add free membranes in
$11$ dimensions, which are mapped onto rational curves in the Calabi--Yau
threefold.  We will discuss this point in the next section.

%%%%%%%%%%%%%%%%%%%%%%%%%%%%%%%%%%%%%%%%%%%%%%%%%%%%%%%%%%%%%%%%%%%%%%

\section{The microscopic picture \label{micr}}

\setcounter{equation}{0}

It has, for a long time, been an open question how to give the
Bekenstein-Hawking entropy a statistical interpretation in terms of a
degeneracy of states. Although there
has been substantial progress 
in terms of the $D$-brane picture \cite{Microscopic}, it is still a
question that deserves further study.

Consider, for example, the special configuration (iii) which,
upon compactification on a Calabi--Yau threefold, yields 
a string in five dimensions.
Inspired by the degeneracy of fundamental strings and the 
``correspondence principle'' \cite{Pol/Hor}, one could argue that the
degeneracy of states of the corresponding (unknown) 
underlying quantum theory should, for large level $N$, be of the form
\begin{eqnarray}
\label{dn}
d(N) &\sim& N^{-\gamma/4} \ e^{2 \pi \sqrt{\frac{c}{6} N}}.
\end{eqnarray}
Here $c$ and $\gamma$ are a priori unknown parameters. For the case 
$\gamma = c + 3$,  
eq. (\ref{dn}) describes the degeneracy of a fundamental string 
with central charge $c$
for large
level $N$ \cite{GSW}.  

The exponential term in  (\ref{dn})
is known as the leading term and the polynomial term as the subleading term.
The leading term is well understood in the context of classical solutions
of supersymmetric vacua, especially for the BPS saturated case
\cite{la/wi,FerKal1,D4N4Entr,BCDKLM}.
The subleading term has been recently identified in the context
of $N=2$ supersymmetric heterotic and type II vacua \cite{be/ga}.
These subleading corrections occur naturally as quantum or instanton 
corrections for
$N < 3$. However, the configuration (iii) is special, and other configurations
(and compactifications) such as, for instance, the case (ii) do
{\em not} share this microscopic picture. 
Moreover, non-extreme black hole entropies in 
effective string theories depend on the values of the moduli at infinity
\cite{cv/ga}. Thus, an interpretation of their entropy
in terms of the degeneracy of the
spectrum of an underlying quantum theory, such as in eq. (\ref{dn}), appears
in general to be somewhat problematic.

In the following we will give a 
microscopic interpretation for certain black hole entropies,
that were derived above in the context of $N=2$ supergravity coupled 
to two vector multiplets, which arises as a low-energy effective string theory.
Such a microscopic interpretation is up to now only possible near 
particular points in moduli space.
In particular  we will propose a microscopic picture that gives a
statistical/thermodynamical interpretation of the 4-d entropy
for the cases (ii) and (iii).

In order to understand the microscopic picture one has to understand
the brane picture. We will first consider case (iii), for which there
is a clear brane picture.  As mentioned in the previous section, the
11-d configuration consists of an intersection of 5-branes and a gas
of closed membranes. The intersection part has been discussed in the
last section and it give rise to the Bekenstein-Hawking entropy
(\ref{075}). A microscopic picture for this part has been given in
\cite{be/mo}. Following the ideas given there, the microscopic states
can be seen as open membrane states that connect the 5-branes. Since
they are massive as long as they are stretched, they will move to the
common intersection in order to become massless there. Next, one wraps the
5-branes around 4-cycles of the Calabi--Yau threefold and obtains the
black string solution given in (\ref{072}). This string is also the
common intersection of all 5-branes, and the open membranes sitting on
the common intersection appear now as momentum modes for this
string. If one further takes into account that the magnetic charge
$p^{\Lambda}$ can be interpreted as arising when wrapping the 5-branes
$p^{\Lambda}$ times around the $\Lambda$-th 4-cycle, one can identify
the Bekenstein-Hawking entropy (\ref{075}) with the statistical
entropy for the string states of this 5-d black string.

%\smallskip

\noindent
This part of the entropy, associated with the intersection of the
5-branes, gets now corrected by an instanton part. The corresponding
microscopic interpretation was given in \cite{be/ga}. Here we will
extend this interpretation further.  The 11-d origin of the instanton
part in the entropy is of a different nature. Turning on the instanton
corrections means that we consider the 11-d intersection to live in a
gas of closed membranes. When compactifying this configuration, the
worldvolume of the closed membranes are completely mapped into the
internal space. Two of the three worldvolume coordinates are mapped
onto rational curves in the Calabi--Yau threefold and the third one is
again identical with the direction of the 5-d black string. The type
IIA analogon would be, that we first compactify over this string (the
$11$-th direction) and obtain in 10-d 4-branes living in a gas of
closed strings. In the second step of the compactification, the
worldvolumes of these closed strings are mapped onto rational curves
in the Calabi--Yau threefold.

%\medskip

\noindent
Keeping this in mind, there emerges a corresponding thermodynamical
picture.  The 11-d intersection lives in thermal equilibrium with a
gas of free closed membranes.  When they touch a 5-brane, they break
up into open membranes which move to the common
intersection. Eventually, they recombine to escaping closed
membranes. The average number of open membranes on the 5-branes is
counted by $q_0$ \cite{be/mo}.  In a thermodynamical picture a natural
definition of the temperature\footnote{One should keep in mind that in
this picture the temperature has nothing to do with the Hawking
temperature of non-extremal black holes. We only want to give a
statistical/thermodynamical picture of the 11-d configuration.} is
given by the radius of the 5-d black string, i.e.
%\ when we neglect the instanton contribution
$T^2 \sim 1/R_5^2 =
(d_{\Lambda\Delta\Sigma}p^{\Lambda} p^{\Delta} p^{\Sigma})^{1 \over
3}/q_0$ (see eq.\ (\ref{072})). By keeping the magnetic charges at some
generic value, the temperature is directly related to the average
number of states on the intersection, i.e.\ to $q_0$.  There are now
two special cases:

\medskip

a) the zero temperature limit ($R_5 \rightarrow \infty$): In this limit all
K\"ahler class moduli are large and hence all instanton corrections
are suppressed.  The black hole states are given by the open membranes
living on the common intersection. Or in the thermodynamical language,
all membranes are condensed - there are no free membranes.
% In field theory this corresponds to the weak coupling limit. Note
% that in the type II string theory picture the radius $R_5$ corresponds
% to $R_{11}$.

b) the infinite temperature limit ($R_5 \ll 1$): There are no open
membranes on the 5-branes (the K\"ahler class moduli are small).  In
this case the instanton corrections yield the dominant part 
%(see eq. (\ref{entzeq})). 
and the black hole states consist of a ``hot gas'' of closed
membranes, which are mapped into the internal space.  The total number
of these states is related to the sum over all rational curves. Note
that this sum is in general infinite. On the other hand the charges
are bounded from below by the brane tension or the zero point
oscillations.  Equivalently the temperature is bounded from above. And
any non-vanishing value of the charges will regularize the instanton sum.  
% In field theory this corresponds to the strong coupling limit.

\medskip

\noindent
In this picture, the transition between the two cases is smooth.  The
reason for this is that we have, so far, implicitly assumed that the
magnetic charges take some finite value.  As a consequence we were
able to change the values of all K\"ahler class moduli in the same
way, i.e.\ we went up and down in the K\"ahler cone. Going down in the
K\"ahler cone means that we heat up the system, which takes us into
the instantonic region.  By going up in the K\"ahler cone, on the
other hand, we cool down the system - all open membranes condense and
we are in the intersection region.

This situation changes, however, drastically if we allow that also one
of the magnetic charges becomes very small. In this case we are
approaching a wall of the K\"ahler cone ($t_1
\rightarrow 0$), where one of the 4-cycles
vanish. At this point, the system undergoes a phase transition, a vanishing
4-cycle ``is replaced'' by an emerging 3-cycle beyond the wall. We do
not wish to discuss this phase transition in detail here.  But if we approach
this point, which on the heterotic side
corresponds to $S \simeq T$,
the entropy gets logarithmic corrections as given in eq.\ (\ref{entrosbt}).

\noindent

Finally, let us discuss a problem related to the $\zeta(3)$ terms
in the prepotential (\ref{preptype2}). As a consequence also
the entropy contains terms wich are proportional to 
$\zeta(3)$. Since this irrational
number cannot be expressed in terms of rational numbers or factors of
$\pi$, it seems to be difficult 
to give the entropy 
a statistical interpretation. In order to address this question 
we can go to a region
in moduli space
where only these terms contribute. This is shown in eq.\ (\ref{entzep}),
where we
took $p^A \simeq 0 $ and $p^0 \gg 1$. The 11-d starting point for this
limit is an intersection of two 2-branes embedded in a gas of closed
membranes. This is a configuration dual to the 5-brane case discussed
above.  As before, in this picture we have open membranes sitting on the
intersection. The contributions proportional to 
$\zeta(3)$ can now be extracted if we go to the hot
temperature limit, i.e.\  $R_5 \ll 1$, which in this case correspond to 
$p^0 \gg 1$. Again the dominant part is given by
the gas of closed membranes. This pure instantonic part yields the
entropy contribution (\ref{entzep}),
and we have to face the problem of interpreting $\zeta(3)$.
Interestingly, this term also appears in the statistical entropy that
counts the number of free bosons and fermions (ideal gas)  
living in the membrane worldvolume, which is given by \cite{kl/ts}
\beqa
S_{stat} = {7 \over 8 \pi} \zeta(3) N L^2 T^2 \;\;\;,
\eeqa
where $N$ denotes the number of states of free bosons that should be
equal to the number of fermions, $L^2$ is the spacial volume (which
should be normalized properly) and $T^2$ is the membrane tension.  The
membrane tension is related to the string tension by \cite{kl/ts}:
$T^2 = T^1 /L = 1/ (2 \pi \alpha' L)$. So, by comparing this
statistical entropy with (\ref{entzep}) and by setting $\alpha'=L$, we
see that both expressions coincide up to integers. This coincidence
suggests, that the $\zeta(3)$ terms in the entropy (\ref{entzep})
count the number of worldvolume states of the compactified $M$-2-branes.

%%%%%%%%%%%%%%%%%%%%%%%%%%%%%%%%%%%%%%%%%%%%%%%%

\section{Conclusions}

In this paper we investigated axion-free quantum black hole solutions in
the $N=2$ supersymmetric heterotic $S$-$T$ model. 
For these solutions we discussed the entropy in target--space
duality invariant form as
well as the scalar fields on the horizon.
% [ I WOULD KILL THIS, i.e.\ $\Re z^A =0$.]     
The entropy in this model is determined by 4 independent charges.  If we
keep all 4 charges the entropy is given by eq.\ (\ref{410}).
This result takes into account {\em all} the perturbative corrections 
appearing in the prepotential. Next, we considered two special
classes of solutions, whose entropy is given in terms of 3 charges only.  
In 11
dimensions, these two cases correspond to intersections of only
membranes or only 5-branes. For the first case the entropy is given in
(\ref{perbh}) and for the second one in (\ref{425}).  In
the latter case we also included the non-perturbative corrections.
However, this latter solution depends on a constrained parameter. We expanded
this solution around a vanishing 4-cycle and found logarithmic
corrections for the entropy (subleading terms). 

In the second part we considered the corresponding 10-d (heterotic) or 11-d
($M$-theory) configurations. In the context of $M$-theory we proposed a
microscopic interpretation for the entropy formulae. In this picture
we have in 11 dimensions an intersection of two branes living in a
gas of free closed membranes. When compactifying this configuration
the intersecting branes are wrapped around inequivalent cycles and the
free closed membranes are mapped onto rational curves of the
CY-threefold. An interesting feature of this model is that, although
the 11-d solution is singular, the compactification on a
CY-threefold stabilizes this solution. A torus compactification, on the
other hand, yields a singular configuration in 4 dimensions.

Finally, we discussed a thermodynamical picture for the intersection of
branes living in a gas of membranes.  The number of open membranes
attached to the intersection of 2- or 5-branes depends on the point in
the K\"ahler cone.  If we move up in the K\"ahler cone, the number
increases (the closed membranes ``condense'') and going down has the
consequence that all open membranes ``evaporate'' from the
intersection. Deep inside the cone we reach a pure instantonic region.
Here, for the case of intersecting membranes, we proposed a
microscopic interpretation for the $\zeta(3)$ terms in the entropy in
terms of worldvolume states of membranes.

To conclude, the microscopic picture of the quantum black hole
solutions we have investigated is not yet complete.  But it is
encouraging that, at least at special points in moduli space, a
reasonable statistical interpretation of the entropy, including
quantum corrections, is possible.

\bigskip  \bigskip

\noindent
{\bf Acknowledgement}  \medskip \newline
We would like to thank E.\ Derrick, M.\ Henningson and T.\ Mohaupt 
for many fruitful discussions and comments.
The work of K.B.~is  supported by the DFG and the work of I.G.
~is supported by U.S. DOE Grant Nos. DOE-EY-76-02-3071 and the National 
Science Foundation Career Advancement Award No. PHY95-12732.
I.G. would like to thank the Humboldt University for hospitality.

%%%%%%%%%%%%%%%%%%%%%%%%%%%%%%%%%%%%%%%%%%%%%%%%%%%%%%%%%%%%%%

%%%%%%%%%%%%%%%%%%%%%%%%%%%%%%%%%%%%%%%%%%%%%%%%%%%%%%%%%%%%%%%%

\end{document}